# A Comparative Study on the Performance of Permutation Algorithms

Youssef Bassil

LACSC – Lebanese Association for Computational Sciences
Registered under No. 957, 2011, Beirut, Lebanon

**Abstract**

*Permutation is the different arrangements that can be made with a given number of things taking some or all of them at a time. The notation P(n,r) is used to denote the number of permutations of n things taken r at a time. Permutation is used in various fields such as mathematics, group theory, statistics, and computing, to solve several combinatorial problems such as the job assignment problem and the traveling salesman problem. In effect, permutation algorithms have been studied and experimented for many years now. Bottom-Up, Lexicography, and Johnson-Trotter are three of the most popular permutation algorithms that emerged during the past decades. In this paper, we are implementing three of the most eminent permutation algorithms, they are respectively: Bottom-Up, Lexicography, and Johnson-Trotter algorithms. The implementation of each algorithm will be carried out using two different approaches: brute-force and divide and conquer. The algorithms codes will be tested using a computer simulation tool to measure and evaluate the execution time between the different implementations.*

## 1. The Permutation Algorithms

The permutation algorithms to be implemented are Bottom-Up [1], Lexicography [2], and Johnson-Trotter [3, 4] algorithms. Each one of them will be implemented using two different approaches: brute-force [5] and divide and conquer [6].

### 1.1. Bottom-Up Algorithm

The bottom-up algorithm starts by taking the given $(n–1)!$ permutations of $\{1,…,n–1\}$, then inserts $n$ into each position of each of them. This generates $n!$ permutations of $\{1,…,n\}$ and they are as follows:

```
Base step:                    1
Step 1:                     1 2, 2 1
Step 2:     1 2 3, 1 3 2, 3 1 2;      3 2 1, 2 3 1, 2 1 3
Step 3:  1234, 1243, 1423, 4123;   4132, 1432, 1342, 1324;
         3124, 3142, 3412, 4312;   4321, 3421, 3241, 3214;
         2314, 2341, 2431, 4231;   4213, 2413, 2143, 2134
```

This ordering is called *minimal change* since each permutation can be obtained from its immediate predecessor by exchanging just two elements in it.

#### 1.1.1. The Brute-Force Pseudo-Code

```
// ALGORITHM BottomUp (input[n])
// Performs permutation using Bottom-Up technique
// INPUT : input (array of integers)
// OUTPUT : list (array of integers)

list: array of integers that holds permutation
index: an integer that points to the current element in the list
input: array of integers that holds the user input
n: an integer that holds the input length
temp: integer used for swapping purposes
current: array of integers holding current permutation instance
m: an integer that holds the current length
```

```
BEGIN
    index ← 0

    List[index] ← input[0]  // adding the 1st permutation instance
    index ← index+1

    k ← 0;
    counter ← 0;

    for (i ← 1 to n-1)
    {
       while(counter<fac(i)) do
       {
          current ← input[i] + list[k]
          list[index] ← current
          index ← index+1

          for (swap ← 0 to m-1) // swap number i with number i+1
          {
             temp ← current[swap + 1]
             current[swap + 1] ← current[swap]
             current[swap] ← temp

             list[index] ← current // add generated permutation
             index ← index+1
             swap ← swap+1
          }

          k ← k+1
          counter ← counter+1
       }
       counter ← 0
       i ← i+1
    }
    RETURN list
END
```

### 1.1.2. The Algorithm Complexity

The algorithm contains three nested loops (*outer for, while, and inner for*) whose bodies are executed consecutively *n-1* times where *n* is the total length of the input, *i!-1* where *i* is the index representing the iteration in which a particular permutation instance is being calculated, and *m-1* where *m* is the total length of the current permutation. Ignoring the instructions outside the loop and taking into consideration the most costly instruction as the basic operation, we get the following:

$$\sum_{i=1}^{n-1} \sum_{counter=0}^{i!-1} \sum_{swap=0}^{m-1} 1 = (n-1)(n-1)! = n!$$ and thus the algorithm is of time complexity $O(n!)$

Since the basic operation is executed *n!* times regardless of the value of the input, we get $C_{Best}(n!) = C_{Worst}(n!) = C_{Average}(n!) = n!$

The complexity of this algorithm is $O(n!)$ where *n* is the length of input to permute. We have three *loops* and one *basic operation* executed *n!* times over *n* elements.

### 1.1.3. The Divide & Conquer Pseudo-Code

// ALGORITHM BottomUpRecursive (input[n])
// Performs Bottom-Up permutation using divide & conquer technique
// INPUT : input (array of integers)
// OUTPUT : list (array of integers)

*list*: array of integers that holds permutation
*index:* an integer that points to the current element in the list *input*: array of integers that holds the user input
*n:* an integer that holds the input length
*i:* an integer representing the current recursive iteration
*temp*: integer used for swapping purposes
*current:* array of integers holding current permutation instance
*m:* an integer that holds the current length

**BEGIN**

    list[index] ← input[0]   // adding the 1st permutation instance
    index←index+1

    BottomUpRecursive(0 , index)

    BottomUpRecursive(i , index)
    {
      k ← 0;
      counter ← 0;

      if (i < n)
      {
        while(counter<fac(i)) do
        {
          current ← input[i] + list[k]
          list[index] ← current
          index←index+1

          for (swap ← 0 to m - 1) // swap number i with number i+1
          {
            temp ←current[swap + 1]
            current[swap + 1] ←current[swap]
            current[swap] ←temp

            list[index] ← current // add generated permutation
            index←index+1

            swap ← swap+1
          }
          k ← k+1
          counter ← counter+1
        }
        counter ← 0
        i ← i+1

        PermuteRecu(i , index) // Recursive Call
      }

    } // end of function

  RETURN list

**END**

### 1.1.4 The Algorithm Complexity

The algorithm contains two nested loops (*while, and inner for*) whose bodies are executed consecutively *i!-1* where *i* is the index representing the iteration in which a particular permutation instance is being calculated, and *m-1* where *m* is the total length of the current permutation. Ignoring the instructions outside the loop and taking into consideration the most costly instruction as the basic operation, we get the following:

$$\sum_{counter=0}^{i!-1} \sum_{swap=0}^{m-1} 1 = m!\text{ and thus the algorithm is of time complexity } O(m!)$$

Since the basic operation is executed *m!* times regardless of the value of the input, we get $C_{Best}(m!)= C_{Worst}(m!)= C_{Average}(m!)= m!$

The complexity of this algorithm is *O(m!)* where *m* is the length of input to permute. We have two *loops* and one *basic operation* executed *m!* times over *m* elements.

## 1.2. Lexicography Algorithm

Given an initial input $p = (p_1, p_2, ..., p_n)$. In order to obtain the next permutation, we must first find the largest index *i* so that $P_i<P_{i+1}$. Then, the element, $P_i$ will be swapped with the smallest of the elements after $P_i$, but not larger than $P_i$. Finally, the last *n - i* elements will be reversed so that they appear in ascending order. This process continues until all permutations are generated.

### 1.2.1. The Brute-Force Pseudo-Code

```
// ALGORITHM Lexicography (input[n])
// Performs permutation using lexicography technique
// INPUT : input (array of integers)
// OUTPUT : list (array of integers)
```

*list*: array of integers that holds permutation
*index:* an integer that points to the current element in the list
*input*: array of integers that holds the user input
*n:* an integer that holds the input length
*temp*: integer used for swapping purposes

**BEGIN**

    index ← 0

    list[index] ← input[0]  // adding the 1st permutation instance
    index←index+1

    for (k ← 0 to < fac(n) - 1) // -1 since one of the permutation is the input which was added earlier
    {
      i ← -1
      j ← 0

      x ← n – 2
      while (x >= 0) do
      {
        if (input[x] < input[x + 1])
        {
          i ← x;
          x←-1 // break from the while loop
        }

        x←x-1
      }

      if (i <> -1)
      {
        x ← n - 1
        while (x > i)
        {
          if (input[x] > input[i])
          {
            j ← x
            x ← I // break from the while loop
          }

```
                x←x-1
            }

        // Swapping elements pointed by i and j;

            temp ←input[i];
            input[i] ←input[j];
            input[j] ←temp;

        // Reversing elements after i28

            Reverse(input, i + 1, n - (i + 1))
        }

    list[index] ← current  // add generated permutation
    index←index+1

    k←k+1
    }

    RETURN list

END
```

### 1.2.2. The Algorithm Complexity

The algorithm contains two nested loops (*outer for, and inner while*) whose bodies are executed consecutively *n!-1* times where *n* is the total length of the input, and *n-2* where *n* is the total length of the input. Ignoring the instructions outside the loop and taking into consideration the most costly instruction as the basic operation, we get the following:

$$\sum_{k=0}^{n!-1} \sum_{x=0}^{n-2} 1 = n-1(n!) = n!$$ and thus the algorithm is of time complexity $O(n!)$

Since the basic operation is executed *n!* times regardless of the value of the input, we get $C_{Best}(n!)= C_{Worst}(n!)= C_{Average}(n!)= n!$

The complexity of this algorithm is $O(n!)$ where *n* is the length of input to permute. We have thwo *loops* and one *basic operation* executed *n!* times over *n* elements.

### 1.2.3. The Divide & Conquer Pseudo-Code

// ALGORITHM LexicographyRecursive (input[n])
// Performs lexicography permutation using divide & conquer technique
// INPUT : input (array of integers)
// OUTPUT : list (array of integers)

*list*: array of integers that holds permutation
*index:* an integer that points to the current element in the list
*input*: array of integers that holds the user input
*n:* an integer that holds the input length
*k:* an integer representing the current recursive iteration
*temp*: integer used for swapping purposes

**BEGIN**

    index ← 0

    List[index] ← input[0]  // adding the 1st permutation instance
    index ← index +1

    LexicographyRecursive(0 , index)

```
LexicographyRecursive (k , index)
{
   i ← -1
   j ← 0

   x ← n – 2
   while (x >= 0) do
   {
      if (input[x] < input[x + 1])
      {
         i ← x;
         x←-1 // break from the while loop
      }

      x←x-1
   }

   if (i <> -1)
   {
      x ← n - 1
      while (x > i)
      {
         if (input[x] > input[i])
         {
            j ← x
            x ← I // break from the while loop
         }

          x←x-1
       }

      // Swapping elements pointed by i and j;

      temp ←input[i];
      input[i] ←input[j];
      input[j] ←temp;

      // Reversing elements after i28

      Reverse(input, i + 1, n - (i + 1))
   }

   list[index] ← current  // add generated permutation
   index←index+1

   k++;
   if (k < fac(n) -1)
   {
        PermuteRecu(k , index) // Recursive Call
    }

} // end of function

RETURN list

END
```

### 1.2.4 The Algorithm Complexity

The algorithm contains one loop (*while*) whose body is executed *n-2* times where *n* is the total length of the input. Additionally, we have a looping recursive call that recursively iterates for *n!-1*. Ignoring the instructions outside the loop and taking into consideration the most costly instruction as the basic operation we get the following:

$$\sum_{i=0}^{n!-1} \sum_{x=0}^{n-2} 1 = (n-1)n! = n! \text{ and thus the algorithm is of time complexity } O(n!)$$

Since the basic operation is executed *n!* times regardless of the value of the input, we get $C_{Best}(n!)= C_{Worst}(n!) = C_{Average}(n!) = n!$

The complexity of this algorithm is *O(n!)* where *m* is the length of input to permute. We have two *loops* and one *basic operation* executed *n*! times over *n* elements.

### 1.3. Johnson-Trotter Algorithm

Generally speaking, the Johnson-Trotter algorithm checks to see whether a mobile number exists or not, if yes the algorithm performs the following:
1. find the largest mobile element *k*
2. swap *k* and the adjacent element it is facing
3. reverse the direction of all elements larger than *k*

As long as there exists a mobile repeat all the above.

### 1.3.1. The Brute-Force Pseudo-Code

```
// ALGORITHM JohnsonTrotter (input[n])
// Performs permutation using Johnson-Trotter
// INPUT : input (array of integers)
// OUTPUT : list (array of integers)
```

*list*: array of integers that holds permutation
*index:* an integer that points to the current element in the list
*input*: array of integers that holds the user input
*n:* an integer that holds the input length
*temp*: integer used for swapping purposes
*pointers*: array of integers that holds present direction of each permutation
*increasingPtr*: array of integers that holds left to right arrows -> -> -> ....
*decreasingPtr*: array of integers that holds right to left arrows <- <- <- ....
*mobile*: integer that holds the mobile element
*mobileIndex*: integer that holds the index of the mobile element
*flag* : boolean variable that indicates if a mobile exists or not
*p:* an integer that holds the pointers array length
*q:* an integer that holds the increasingPtr array length
*r:* an integer that holds the decreasingPtr array length

**BEGIN**

    index ← 0

    list[index] ← input[0]   // adding the 1st permutation instance
    index←index+1

    //Initialize pointers <- <- <- ....
    for (i ← p – 1 to 0)
      pointers[i] ← i - 1
      i←i-1

    //Initialize increasingPtr -> -> -> ....
    for (i ← 0 to q-1)
      increasingPtr[i] ← i + 1
      i←i+1

```
//Initialize decreasingPtr <- <- <- ....
for (i ← r – 1 to 0)
   decreasingPtr[i] ← i – 1
   i←i-1

for (i ← 0 to fac(n) - 1)  // -1 since one of the permutation is the input which was added earlier
{
mobile ← 0
mobileIndex ← 0
flag ← false

//Find the largest Mobile

for (i ← 0 to n-1)
{
    if (pointers[i] <> -1 AND pointers[i] <> n AND input[i] > mobile AND input[pointers[i]] < input[i])
    {
       mobile ← input[i]
       mobileIndex ← i
       flag ← true
    }
}

if (flag = true)
{
   // Swap

   input[mobileIndex] ← input[pointers[mobileIndex]]
   input[pointers[mobileIndex]] ← mobile

   if (pointers[pointers[mobileIndex]] = mobileIndex)
   {
     if (pointers[mobileIndex] > mobileIndex)
     {
       temp ←pointers[pointers[mobileIndex]]
       pointers[pointers[mobileIndex]] ←pointers[mobileIndex] + 1
       pointers[mobileIndex] ←temp - 1
     }
     else
     {
       int temp ← pointers[pointers[mobileIndex]]
       pointers[pointers[mobileIndex]] ← pointers[mobileIndex] - 1
       pointers[mobileIndex] ← temp + 1
     }

   }

}

// Reverse Directions(arrows)

for (int j ← 0 to n-1)
{
   if (input[j] > mobile)
     if (pointers[j] = increasingPtr[j])
        pointers[j] ← decreasingPtr[j]
     else if (pointers[j] = decreasingPtr[j])
        pointers[j] ← increasingPtr[j]

       j←j+1
     }
```

      i←i+1

  list[index] ← current // add generated permutation
  index←index+1

 } // end of outer for loop

 RETURN list

**END**

### 1.3.2. The Algorithm Complexity

The algorithm contains two nested loops (*outer for, and inner for*) whose bodies are executed consecutively *n!-1* times where *n* is the total length of the input, and *n-1* where *n* is the total length of the input. Ignoring the instructions outside the loop and taking into consideration the most costly instruction as the basic operation we get the following:

$$\sum_{k=0}^{n!-1} \sum_{x=0}^{n-1} 1 = n(n!) = n! \text{ and thus the algorithm is of time complexity } O(n!)$$

Since the basic operation is executed *n!* times regardless of the value of the input, we get $C_{Best}(n!)= C_{Worst}(n!)= C_{Average}(n!)= n!$

The complexity of this algorithm is *O(n!)* where *n* is the length of input to permute. We have two *loops* and one *basic operation* executed *n*! times over *n* elements.

### 1.3.3. The Divide & Conquer Pseudo-Code

 // ALGORITHM JohnsonTrotterRecursive (input[n])
 // Performs Johnson-Trotter permutation using divide & conquer technique
 // INPUT : input (array of integers)
 // OUTPUT : list (array of integers)

 *list*: array of integers that holds permutation
 *index:* an integer that points to the current element in the list
 *input*: array of integers that holds the user input
 *n:* an integer that holds the input length
 *temp*: integer used for swapping purposes
 *pointers*: array of integers that holds present direction of each permutation
 *increasingPtr*: array of integers that holds left to right arrows -> -> -> ....
 *decreasingPtr*: array of integers that holds right to left arrows <- <- <- ....
 *mobile*: integer that holds the mobile element
 *mobileIndex*: integer that holds the index of the mobile element
 *flag* : boolean variable that indicates if a mobile exists or not
 *p:* an integer that holds the pointers array length
 *q:* an integer that holds the increasingPtr array length
 *r:* an integer that holds the decreasingPtr array length
 *k:* an integer representing the current recursive iteration

 **BEGIN**

  index ← 0

  List[index] ← input[0] // adding the 1st permutation instance
  index ← index +1

  JohnsonTrotterRecursive (0 , index)

  JohnsonTrotterRecursive (k , index)
  {
   mobile ← 0
   mobileIndex ← 0

```
            flag ← false

            //Find the largest Mobile

            for (i ← 0 to n)
            {
                if (pointers[i] <> -1 AND pointers[i] <> n AND input[i] > mobile AND input[pointers[i]] < input[i])
                {
                    mobile ← input[i]
                    mobileIndex ← i
                    flag ← true
                }
            }

            if (flag = true)
            {
              // Swap

              input[mobileIndex] ← input[pointers[mobileIndex]]
              input[pointers[mobileIndex]] ← mobile

              if (pointers[pointers[mobileIndex]] = mobileIndex)
              {
                 if (pointers[mobileIndex] > mobileIndex)
                 {
                    temp ←pointers[pointers[mobileIndex]]
                    pointers[pointers[mobileIndex]] ←pointers[mobileIndex] + 1
                    pointers[mobileIndex] ←temp - 1
                 }
                 else
                 {
                    int temp ← pointers[pointers[mobileIndex]]
                    pointers[pointers[mobileIndex]] ← pointers[mobileIndex] - 1
                    pointers[mobileIndex] ← temp + 1
                 }

              }

            }

            // Reverse Directions(arrows)

            for (int j ← 0 to n)
            {
              if (input[j] > mobile)
                 if (pointers[j] = increasingPtr[j])
                    pointers[j] ← decreasingPtr[j]
                 else if (pointers[j] = decreasingPtr[j])
                    pointers[j] ← increasingPtr[j]

                  j←j+1
             }

              i←i+1

            list[index] ← current   // add generated permutation
            index ←index+1

         k++;
         if (k < fac(n) -1)
         {
              PermuteRecu(k , index) // Recursive Call
         }
```

   } // end of function

   RETURN list

  **END**

### 1.3.4. The Algorithm Complexity

The algorithm contains one loop (*for*) whose body is executed *n* times where *n* is the total length of the input. Additionally, we have a looping recursive call that recursively iterates for *n!-1*. Ignoring the instructions outside the loop and taking into consideration the most costly instruction as the basic operation, we get the following:

$$\sum_{i=0}^{n!-1} \sum_{x=0}^{n} 1 = (n+1)n! = n!$$ and thus the algorithm is of time complexity $O(n!)$

Since the basic operation is executed *n!* times regardless of the value of the input, we get $C_{Best}(n!)= C_{Worst}(n!)= C_{Average}(n!)= n!$

The complexity of this algorithm is $O(n!)$ where *m* is the length of input to permute. We have two *loops* and one *basic operation* executed *n*! times over *n* elements.

## 2. Implementation

The six permutation algorithms were all implemented using MS C#.NET 2008 [7] under the .NET Framework 3.5 [8] and MS Visual Studio 2008. Figure 1, 2, and 3 are screenshots that depict the different results obtained for the different implementations.

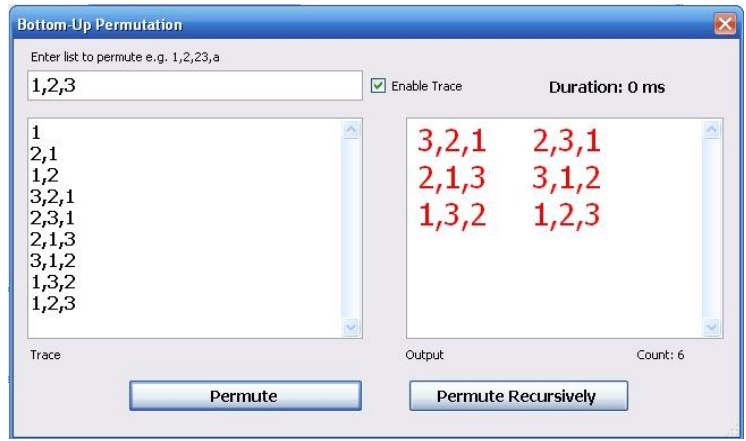

Figure 1 - Bottom-Up Permutation

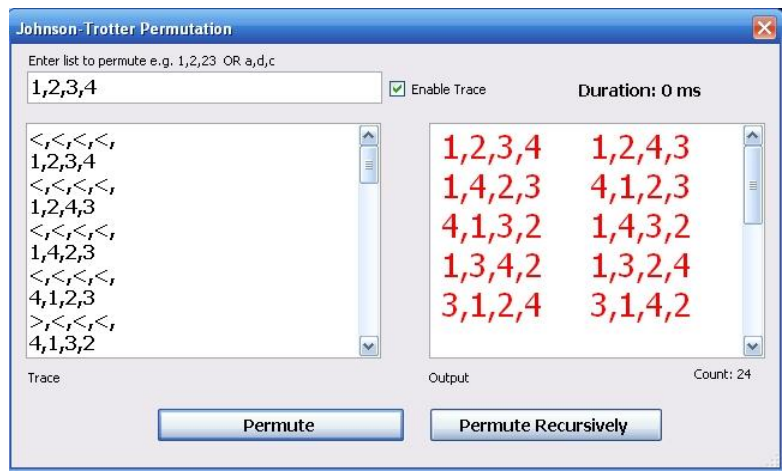

Figure 2 - Johnson-Trotter Permutation

17

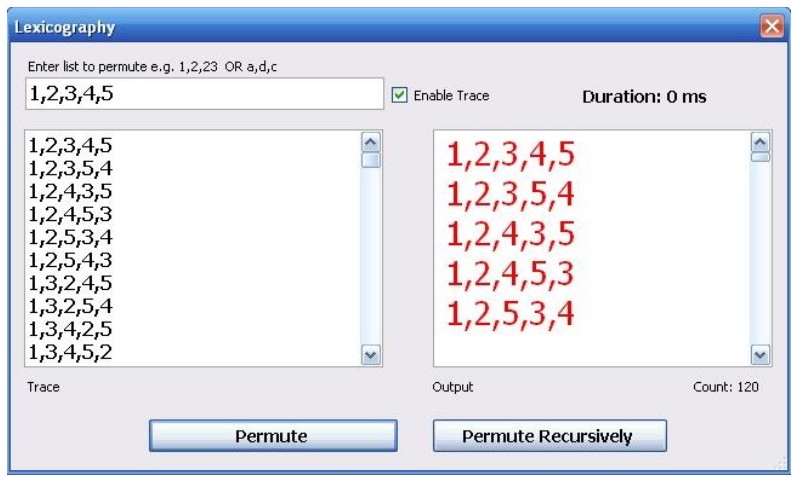

Figure 3 - Lexicography Permutation

## 3. Testing & Experiments

A comparison of the execution time between the six permutation algorithms was undertaken using a desktop IBM-compatible PC with Intel Core 2 dual core processor with 2.66 MHz clock speed, 256KB of cache, and 2GB of RAM. The operating system used was MS Windows Vista. It is worth noting that the execution time for all different algorithms is the average time obtained after five consecutive runs of the same test. Table 1 delineates the execution time of the permutation algorithms using the brute-force method; while, Table 2 using the divide and conquer method.

Table 1 - Results obtained for the three brute-force permutation algorithms

| Test Case | Input Length | Bottom-Up | Lexicography | Johnson-Trotter |
|---|---|---|---|---|
| 1 | 4 | < 1ms | < 1ms | < 1ms |
| 2 | 6 | < 1ms | < 1ms | < 1ms |
| 3 | 8 | 187.5 ms | 328 ms | 46 ms |
| 4 | 9 | 1.45 s | 2.1 s | 437 ms |
| 5 | 10 | 14.9 s | 20.4 s | 3.52 s |

Table 2 - Results obtained for the three divide & conquer permutation algorithms

| Test Case | Input Length | Bottom-Up | Lexicography | Johnson-Trotter |
|---|---|---|---|---|
| 1 | 4 | < 1ms | < 1ms | < 1ms |
| 2 | 6 | < 1ms | < 1ms | < 1ms |
| 3 | 8 | 171.2 ms | 301 ms | 47.11 ms |
| 4 | 9 | 1.79 s | 2.02 s | 469 ms |
| 5 | 10 | 16.7 s | 22.7 s | 3.2 s |

## 4. Conclusions

From the obtained results delineated in tables 1 and 2, it is obvious that the Johnson-Trotter permutation algorithm outsmarted all other algorithms in all different test cases. When input lengths were respectively 4, 6, and 8 in length, the six algorithms performed equally. However, when the length became as large as 9, the Johnson-Trotter algorithm surpassed the others by around 1.5 seconds. Additionally, the Johnson-Trotter algorithm showed impressive results as compared to others when the input length reached 10. It then surpassed the other algorithms by around 14 seconds.

## Acknowledgment

This research was funded by the Lebanese Association for Computational Sciences (LACSC), Beirut, Lebanon, under the "Evaluation & Performance Research Project – EPRP2012".